\documentclass[preprintnumbers, floatfix,letterpaper,aps,prd,epsfig,nofootinbib,
twocolumn
]{revtex4-1}
\usepackage{bm,graphicx,dcolumn,epstopdf,epsf, latexsym,mathbbol, amssymb,amsmath,color,slashed, mathrsfs,mathcomp,simplewick}
\usepackage[center]{subfigure}
\usepackage{multirow}
\usepackage{makecell}
\usepackage[colorlinks,linkcolor=blue,citecolor=blue,urlcolor=blue]{hyperref}

\begin{document}
\allowdisplaybreaks
 \newcommand{\bq}{\begin{equation}}
 \newcommand{\eq}{\end{equation}}
 \newcommand{\bqn}{\begin{eqnarray}}
 \newcommand{\eqn}{\end{eqnarray}}
 \newcommand{\nb}{\nonumber}
 \newcommand{\lb}{\label}
 \newcommand{\f}{\frac}
 \newcommand{\p}{\partial}
\newcommand{\PRL}{Phys. Rev. Lett.}
\newcommand{\PLB}{Phys. Lett. B}
\newcommand{\PRD}{Phys. Rev. D}
\newcommand{\CQG}{Class. Quantum Grav.}
\newcommand{\JCAP}{J. Cosmol. Astropart. Phys.}
\newcommand{\JHEP}{J. High. Energy. Phys.}
\newcommand{\red}{\textcolor{black}}

\title{Polarized primordial gravitational waves in the ghost-free parity-violating gravity}

\author{Jin Qiao${}^{a, b}$}
\email{qiaojin@zjut.edu.cn}

\author{Tao Zhu${}^{a, b}$}
\email{zhut05@zjut.edu.cn}

\author{Wen Zhao${}^{c, d}$}
\email{wzhao7@ustc.edu.cn; Corresponding author}

 \author{Anzhong Wang${}^{e}$}
 \email{anzhong$\_$wang@baylor.edu}

\affiliation{${}^{a}$ Institute for theoretical physics and Cosmology, Zhejiang University of Technology, Hangzhou, 310032, China\\
${}^{b}$ United center for gravitational wave physics (UCGWP), Zhejiang University of Technology, Hangzhou, 310032, China \\
${}^{c}$ CAS Key Laboratory for Research in Galaxies and Cosmology, Department of Astronomy, University of Science and Technology of China, Hefei 230026, China; \\
${}^{d}$ School of Astronomy and Space Sciences, University of Science and Technology of China, Hefei, 230026, China\\
${}^{e}$ GCAP-CASPER, Physics Department, Baylor University, Waco, TX 76798-7316, USA}

\date{\today}

%
%

\begin{abstract}
The tests of parity symmetry in the gravitational interaction is an attractive issue in the gravitational-wave astronomy. In the general theories of gravity with parity violation, one of the fundamental results is that the primordial gravitational waves (PGWs) produced during the slow-roll inflation is circularly polarized. In this article, we investigate the polarization of PGWs in the recently proposed ghost-free parity-violating gravity, which generalizes the Chern-Simons gravity by including the higher derivatives of the coupling scalar field. For this purpose, we first construct the approximate analytical solution to the mode function of the PGWs during the slow-roll inflation by using the uniform asymptotic approximation. With the approximate solution, we calculate explicitly the power spectrum and the corresponding circular polarization of the PGWs analytically, and find that the contributions of the higher derivatives of the coupling scalar field to the circular polarization is of the same order of magnitude as that of Chern-Simons gravity. The degree of circular polarization of PGWs is suppressed by the energy scale of parity violation in gravity, which is hardly detected on the basis solely of two-point statistics from future cosmic microwave background data.

\end{abstract}


\maketitle

\section{Introduction}
\renewcommand{\theequation}{1.\arabic{equation}} \setcounter{equation}{0}

The precise measurements of the cosmic microwave background (CMB), in particular by WMAP and Planck missions, provide the invaluable information about the physics in the very early universe. Their angular power spectra directly reflect statistical properties of primordial density fluctuations and gravitational waves (PGWs) \cite{Gorski:1996cf, WMAP, Aghanim:2018eyx, Akrami:2018odb}, which are in good agreement with the theoretical predictions of the slow-roll inflation model with a single scalar field \cite{inflation1, inflation2, inflation3, inflation4, inflation5, inflation6}.  For a standard slow-roll inflation in the framework of general relativity (GR), the PGWs have two polarization modes which share exactly the same statistical properties and their primordial power spectra take the same form. Such PGWs produce the TT, EE, BB, and TE spectra of CMB, but the TB and EB spectra vanish because of the parity symmetry of GR \cite{TTEEBB1, TTEEBB2,TTEEBB3,TTEEBB4,TTEEBB}. Since nonzero TB and EB spectra of CMB implies parity violation in the gravitational sector,  the precise measurement of TB and EB spectra could be an important evidences of the parity violation of the gravitational interaction \cite{parity_CMB, parity_CMB1, parity_CMB2, parity_CMB3, parity_CMB4} \footnote{Note that, another effect to produce the TB and EB power spectra of CMB is the so-called cosmological birefringence effect, which can be caused by the possible coupling between the electromagnetic field and the scalar
field through the Chern-Simons term \cite{parity_CMB,CPT1,CPT3,CPT4,CPT5}. }.

In fact, the gravitational terms with parity violation are ubiquitous in numerous candidates of quantum gravity, such as string theory, loop quantum gravity, and Horava-Lifshitz gravity. One well-studied example is the gravitational Chern-Simons term, which arises from string theory \cite{parity_CMB, parity_CMB1, parity_CMB2, parity_CMB3, parity_CMB4, parity_string, parity_string01, parity_string02, parity_string2} and loop quantum gravity \cite{LQG, LQG01, LQG02, LQG03, LQG04}. In Horava-Lifshitz gravity, the parity-violating third and fifth spatial derivative terms are allowed in the gravitational action of the theory \cite{parity_power8, Zhu:2013fja, Takahashi:2009wc}. In the literature, parity violation can also arises from graviton self-couplings \cite{1011, 1011a}, gaugeflation and chromonatural inflation \cite{1215, 1215a, 1215b, 1215c}, Holst gravity \cite{16}, and in models that connect leptogenesis to PGWs \cite{parity_string01, 1718a}. In all these examples, a fundamental effect of the parity violation is the circular polarization of the PGWs, i.e., the left-hand and right-hand polarization modes of GWs propagate with different behaviors. As we mentioned above, such asymmetry between two chiral modes of PGWs can induce a significant parity-violating signature in the CMB polarization (E/B) power spectra, which has motivated a lot of works in this direction (see \cite{parity_power, parity_power1, parity_power2, parity_CMB4, parity_power4, parity_power5, parity_power6, parity_power7, parity_power8, parity_power9, parity_power10, Cai:2016ihp} and references therein for examples).

Recently, based on the Chern-Simons modified gravity, a ghost-free parity-violating theory of gravity has been explored in \cite{ghost_free} by including the higher derivatives of the coupling scalar field. The observational implications of this theory, as well as its extensions of the gravitational waves generated by the compact binaries, have been explored in a series papers \cite{japan,Gao_parity_2019, Zhao_Pgw_2018, Zhu_gravity_2019}. In comparison with the Chern-Simons gravity, one of the distinguishable features of the higher derivatives of the coupling scalar field is that it leads to the velocity birefringence phenomenon, i.e., the velocities of left-hand and right-hand circular polarizations of GWs are different in the ghost-free parity-violating gravities. Thus, it is expected that such velocity birefringence effect could induce some distinguishable signatures in the power spectrum PGWs. With these motivations, in this paper we study the circularly polarized PGWs in this theory of gravity with parity violation, and the possibility to detect the chirality of PGWs by future potential CMB observations.

The rest of the paper is organized as follows: In Sec.~II we give a very brief review on the ghost-free parity-violating gravities, and in Sec.~III we consider a flat Friedmann-Robertson-Walker (FRW) universe and derive the equation of motion for the PGWs. In Sec.~IV,  we first construct the approximate analytical solution to the PGWs by using the uniform asymptotic approximation, and then calculate explicitly the power spectrum and the polarization of PGWs during the slow-roll inflation. The effects of the parity violation in the CMB spectra and their detectability have also been briefly discussed. The paper is ended with Sec.~V, in which we summarize our main conclusions and provide some outlooks.

\section{Parity-violating gravities}
\renewcommand{\theequation}{2.\arabic{equation}} \setcounter{equation}{0}

In this section, we present a brief introduction of the ghost-free parity-violating gravity proposed in \cite{ghost_free}. The action of parity-violating gravity has the form
\bqn\lb{action}
S &=& \frac{1}{16\pi G} \int d^4 x \sqrt{-g}(R+\mathcal{L}_{\rm PV}+\mathcal{L}_{\phi})   ,
\eqn
where $R$ is the Ricci scalar, $\mathcal{L}_{\rm PV}$ is a parity-violating Lagrangian, and $\mathcal{L}_\phi$ is the Lagrangian for scalar field, which is coupled non-minimally to gravity. As a simplest example, we consider the action of the scalar field as
\bqn
\mathcal{L}_\phi =   \frac{1}{2} g^{\mu \nu} \partial_\mu \phi \partial_\nu \phi +V(\phi).
\eqn
Here $V(\phi)$ denotes the potential of the scalar field. The parity-violating Lagrangian of the theory can be written in the form
\bqn
\mathcal{L}_{\rm PV} = \mathcal{L}_{\rm CS} + \mathcal{L}_{\rm PV1} + \mathcal{L}_{\rm PV2},
\eqn
where the Chern-Simons term $\mathcal{L}_{\rm CS}$ is given by
\bqn
\mathcal{L}_{\rm CS} = \frac{1}{8}\vartheta(\phi) \varepsilon^{\mu\nu\rho\sigma} R_{\rho\sigma \alpha\beta} R^{\alpha \beta}_{\;\;\;\; \mu\nu},
\eqn
with $\varepsilon_{\rho \sigma \alpha \beta}$ the Levi-Civit\'{a} tensor defined in terms of the the antisymmetric symbol $\epsilon^{\rho \sigma \alpha \beta}$ as $\varepsilon^{\rho \sigma \alpha \beta}=\epsilon^{\rho \sigma \alpha \beta}/\sqrt{-g}$. $\mathcal{L}_{\rm PV1}$ contains the first derivative of the scalar field and is given by
\bqn
\mathcal{L}_{\rm PV1} &=& \sum_{\mathcal{A}=1}^4  a_{\mathcal{A}}(\phi, \phi^\mu \phi_\mu) L_{\mathcal{A}},\\
L_1 &=& \varepsilon^{\mu\nu\alpha \beta} R_{\alpha \beta \rho \sigma} R_{\mu \nu\; \lambda}^{\; \; \;\rho} \phi^\sigma \phi^\lambda,\nonumber\\
L_2 &=&  \varepsilon^{\mu\nu\alpha \beta} R_{\alpha \beta \rho \sigma} R_{\mu \lambda }^{\; \; \;\rho \sigma} \phi_\nu \phi^\lambda,\nonumber\\
L_3 &=& \varepsilon^{\mu\nu\alpha \beta} R_{\alpha \beta \rho \sigma} R^{\sigma}_{\;\; \nu} \phi^\rho \phi_\mu,\nonumber\\
L_4 &=&  \varepsilon^{\mu\nu\rho\sigma} R_{\rho\sigma \alpha\beta} R^{\alpha \beta}_{\;\;\;\; \mu\nu} \phi^\lambda \phi_\lambda,\nonumber
\eqn
with $\phi^\mu \equiv \nabla^\mu \phi$. It has been shown that in order to avoid the Ostrogradsky modes, it is required that $4a_1+2 a_2+a_3 +8 a_4=0$. The term $\mathcal{L}_{\rm PV2}$, which contains the second derivatives of the scalar field, is described by
\bqn
\mathcal{L}_{\rm PV2} &=& \sum_{\mathcal{A}=1}^7 b_{\mathcal{A}} (\phi,\phi^\lambda \phi_\lambda) M_{\mathcal{A}},\\
M_1 &=& \varepsilon^{\mu\nu \alpha \beta} R_{\alpha \beta \rho\sigma} \phi^\rho \phi_\mu \phi^\sigma_\nu,\nonumber\\
M_2 &=& \varepsilon^{\mu\nu \alpha \beta} R_{\alpha \beta \rho\sigma} \phi^\rho_\mu \phi^\sigma_\nu, \nonumber\\
M_3 &=& \varepsilon^{\mu\nu \alpha \beta} R_{\alpha \beta \rho\sigma} \phi^\sigma \phi^\rho_\mu \phi^\lambda_\nu \phi_\lambda, \nonumber\\
M_4 &=& \varepsilon^{\mu\nu \alpha \beta} R_{\alpha \beta \rho\sigma} \phi_\nu \phi_\mu^\rho \phi^\sigma_\lambda \phi^\lambda, \nonumber\\
M_5 &=& \varepsilon^{\mu\nu \alpha \beta} R_{\alpha \rho\sigma \lambda } \phi^\rho \phi_\beta \phi^\sigma_\mu \phi^\lambda_\nu, \nonumber\\
M_6 &=& \varepsilon^{\mu\nu \alpha \beta} R_{\beta \gamma} \phi_\alpha \phi^\gamma_\mu \phi^\lambda_\nu \phi^\lambda, \nonumber\\
M_7 &=& (\nabla^2 \phi) L_1,\nonumber
\eqn
with $\phi^{\sigma}_\nu \equiv \nabla^\sigma \nabla_\nu \phi$. Similarly, in order to avoid the Ostrogradsky modes in the unitary gauge, the following conditions should be imposed: $b_7 = 0$, $b_6 = 2(b_4 + b_5)$ and $b_2 =-A_*^2(b_3 -b_4)/2$, where $A_*\equiv \dot \phi(t)/N$ and $N$ is the lapse function of the spacetime.

In the current paper, we focus only on the terms coupled with the first and second derivatives of the scalar field. The more general forms of the Lagrangian, which contains higher-order derivatives of the scalar field, can be found in \cite{Gao_parity_2019}.

\section{Gravitational wave in parity-violation gravities}
\renewcommand{\theequation}{3.\arabic{equation}} \setcounter{equation}{0}

In the flat FRW universe, the background metric is given by
\bqn
ds^2 = a^2(\tau) (-d\tau^2 + \delta_{ij}dx^i dx^j),
\eqn
where $a(\tau)$ denotes the scale factor of the universe and $\tau$ represents the conformal time, which relates to the cosmic time $t$ via $dt = a(\tau) d\tau$. In the parity-violating gravities considered in this paper, we observe that all the parity-violating terms in the action (\ref{action}) have no effect on the background evolution. We further assume that the universe is dominated by the scalar field $\phi$ which plays the role of the inflaton field during the slow-roll inflation. In this case, the Friedmann equation, which governs the background evolution, takes exactly the same form as that in GR, i.e.,
\bqn
H^2 = \frac{8 \pi G}{3} \rho,
\eqn
where $H$ denotes the Hubble parameter during the inflationary stage, and the energy density of scalar field is $\rho=\frac{1}{2}\dot \phi^2 +V(\phi)$. The evolution of the scalar field $\phi$ is also the same as that in GR,
\bqn
\ddot \phi + 3 H \dot \phi + \frac{dV(\phi)}{d\phi}=0.
\eqn
It is worth noting that in the standard slow-roll inflation, the scalar field is assumed to satisfy the slow-roll conditions,
\bqn\lb{Con}
|\ddot \phi| \ll |3 H \dot \phi|, \;\;\; |\dot\phi^2| \ll V(\phi).
\eqn
With the above slow-roll conditions, it is convenient to define the following Hubble slow-roll parameters,
\bqn
\epsilon_1 = - \frac{\dot H}{H^2},\;\; \epsilon_2 = \frac{d \ln \epsilon_1}{d \ln a}, \;\; \epsilon_3 =  \frac{d \ln \epsilon_2}{d \ln a}.
\eqn

Primordial gravitational waves are the tensor perturbations of the homogeneous and isotropic background, and we turn to study their propagation. With the tensor perturbations, the spatial metric is written as
\bqn
g_{ij}=a^2(\tau) (\delta_{ij} + h_{ij}(\tau, x^i)),
\eqn
where $h_{ij}$ represents the transverse and traceless metric perturbations, i.e.,
\bqn
\partial^i h_{ij} =0 = h_i^i.
\eqn
In order to derive the equation of motion for the tensor perturbations, we substitute the metric perturbation into the action (\ref{action}) and expand it to the second order in $h_{ij}$. After tedious calculations, we find
\bqn
S^{(2)} = \frac{1}{16\pi G} \int d\tau d^3 x a^4(\tau) \left[ \mathcal{L}_{\rm GR}^{(2)} + \mathcal{L}_{\rm PV}^{(2)}\right],
\eqn
where
\bqn
\mathcal{L}_{\rm GR}^{(2)} &=& \frac{1}{4 a^2} \left[ (h'_{ij})^2 - (\partial_k h_{ij})^2\right],\\
\mathcal{L}_{\rm PV}^{(2)} &=& \frac{1}{4 a^2} \left[\frac{c_1(\tau)}{aM_{\rm PV}} \epsilon^{ijk}h_{il}' \partial_j h_{kl}'+ \frac{c_2(\tau)}{a M_{\rm PV} }   \epsilon^{ijk}\partial^2h_{il}  \partial_j h_{kl}\right], \nb\\
\eqn
where $M_{\rm PV}$ denotes the characteristic energy scale of the parity violation in the theory.

In the above expression, $c_1$ and $c_2$ represent the dimensionless coefficients normalized by the characteristic energy scale $M_{\rm PV}$, which are given by \cite{Zhu_gravity_2019}
\bqn
\frac{ c_1(\tau)}{M_{\rm PV}} &=& \dot{\vartheta}-4 \dot{a_1}\dot{\phi}^2 -8 a_1\dot{\phi}\ddot{\phi} + 8a_1 H \dot{\phi}^2- 2\dot{a_2}\dot{\phi}^2 - 4a_2\dot{\phi}\ddot{\phi}\nb\\
&&+\dot{a_3}\dot{\phi}^2 +2a_3\dot{\phi}\ddot{\phi} -4a_3 H \dot{\phi}^2-4\dot{a_4}\dot{\phi}^2 -8a_4\dot{\phi}\ddot{\phi}\nb\\
&&-2 b_1\dot{\phi}^3+4b_2\left(2 H\dot{\phi}^2-\dot{\phi}\ddot{\phi}\right)\nb\\
&&+2b_3\left(\dot{\phi}^3\ddot{\phi}- H \dot{\phi}^4\right)+2b_4\left(\dot{\phi}^3\ddot{\phi}- H \dot{\phi}^4\right)\nb\\
&&-2b_5 H \dot{\phi}^4+2b_7\dot{\phi}^3\ddot{\phi},\\
\frac{c_2(\tau)}{M_{\rm PV}} &=& \dot{\vartheta}-2\dot{a_2}\dot{\phi}^2 -4a_2\dot{\phi}\ddot{\phi} -\dot{a_3}\dot{\phi}^2 -2a_3\dot{\phi}\ddot{\phi} \nb\\
&&-4\dot{a_4}\dot{\phi}^2 -8a_4\dot{\phi}\ddot{\phi}.
\eqn
Then varying the action with respect to $h_{ij}$, we obtain the field equation for $h_{ij}$ \cite{Zhu_gravity_2019},
\bqn\label{3.13}
&&h_{ij}'' + 2 \mathcal{H} h_{ij}'  - \partial^2 h_{ij}  \nb\\
&&~+ \frac{\epsilon^{ilk}}{aM_{\rm PV}} \partial_l \Big[ c_1 h_{jk}'' + (\mathcal{H}c_1+c_1') h_{jk}' - c_2 \partial^2 h_{jk}\Big]=0.\nb\\
\eqn

\section{Polarization of PGWs}
\renewcommand{\theequation}{4.\arabic{equation}} \setcounter{equation}{0}

\subsection{Equation of motion for GWs}

In the parity-violating gravities, the propagation equations for the two circular polarization modes of GWs are decoupled. To study the evolution of $h_{ij}$, we expand it over spatial Fourier harmonics,
\bqn
h_{ij}(\tau, x^i) = \sum_{A={\rm R, L}} \int \frac{d^3 k}{(2\pi)^3} \tilde h_A(\tau, k^i)e^{i k_i x^i} e_{ij}^{A}(k^i),\nb\\
\eqn
where $e_{ij}^A$ denote the circular polarization tensors and satisfy the relation
\bqn
\epsilon_{i l m} k^l e_{ij}^A = i k\rho_A e^A_{mj},
\eqn
with $\rho_{\rm R}=1$ and $\rho_{\rm L} =-1$. Thus, the field equation in Eq.(\ref{3.13}) can be casted into the form \cite{Zhu_gravity_2019}
\bqn\label{eq4.3}
\tilde h_A'' + (2+\nu_A) \mathcal{H} \tilde h_A' + (1+\mu_A) k^2 \tilde h_A=0,
\eqn
where a $\emph{prime}$ denotes the derivative with respect to the conformal time $\tau$. The deviations from that in GR are quantified by the quantities $\nu_A$ and $\mu_A$, which are given by
\bqn
\nu_A &=& \frac{\rho_A k (c_1 \mathcal{H} -c_1' )/(a \mathcal{ H}  M_{\rm PV})}{1- \rho_A k c_1/(a M_{\rm PV})}, \lb{nuA}\\
\mu_A &=& \frac{\rho_A k (c_1 -c_2)/(a M_{\rm PV})}{1- \rho_A k c_1/(a M_{\rm PV})}.\lb{muA}
\eqn
The quantity $\nu_A$ describes the modification of the friction term, and $\mu_{A}$ describes the modification of the dispersion relation of GWs. In the parity-violating gravities, the former induces the amplitude birefringence effect of GWs, while the latter induces the velocity birefringence of GWs. In the specific case with $c_1/M_{\rm PV} =c_2/M_{\rm PV} = \dot \vartheta$, this equation reduces to that in Chern-Simons gravity, in which we have $\mu_A=0$, i.e., only the amplitudes of GWs are modified during the propagation through the term $\nu_A$. However, in the general case of ghost-free parity-violating gravity, both $\nu_{A}$ and $\mu_{A}$ are nonzero. In particular, $\mu_A \neq 0$ represents a distinguishable features of the higher derivatives of the coupling scalar field in this theory, which leads to the velocity birefringence phenomenon, i.e., the velocities of left-hand and right-hand modes of GWs are different in the ghost-free parity-violating gravity.

As usual, we define the variable $u_A\equiv zh_A$, and rewrite Eq.(\ref{eq4.3}) as,
\bqn\lb{Eq}
u''_A+\left[(1+\mu_A)k^2-{z''}/{z}\right]u_A=0,
\eqn
where $z=a\left(1-{c_1k\rho_A}/{(aM_{\rm PV})}\right)^{1/2}$. In this article, we consider the PGWs during the inflationary stage, and assume that the background evolution during the inflation is slowly-varying. In addition, we expect the deviations from GR arising from the parity violation are small. With this considerations, we can expand the effective time-dependent mass term $z''/z$ in (\ref{Eq}) in terms of the slow-roll parameters and corrections from the parity violation as
\bqn\lb{z}
\frac{z''}{z}&=&\frac{a''}{a}+\frac{1}{2}\frac{\left(\frac{a''}{a}c_1-c_1''\right)k\rho_A/aM_{\rm PV}}{1-c_1k\rho_A/aM_{\rm PV}}\nb\\
&&+\frac{1}{4}\left[\frac{(c_1\mathcal{H}-c_1')k\rho_A/aM_{\rm PV}}{1-c_1k\rho_A/aM_{\rm PV}}\right]^2\nb\\
&\simeq&\frac{v_t^2-\frac{1}{4}}{\tau^2}-\rho_A\frac{k}{\tau}c_1\epsilon_*,
\eqn
where
\bqn
v_t \simeq \frac{3}{2}+3 \epsilon_1+4\epsilon_1^2+4\epsilon_1\epsilon_2+\mathcal{O}(\epsilon^3),
\eqn
and $\epsilon_* \equiv {H}/{M_{\rm PV}}$ denotes the ratio between the energy scale of inflation and the characteristic energy scale of the parity violation, which determines the magnitude of the corrections to GR. \red{From both the expressions of $z''/z$ in (\ref{z}) and $\mu_A$ in (\ref{nuA}), one observes that there is a divergences if $k_{\rm phys} /M_{\rm PV} \sim 1$ with $k_{\rm phys} \equiv k/a$ and $c_1 \sim \mathcal{O}(1)$ during the slow-roll inflation. As pointed out in \cite{parity_power}, with the existence of this divergence, the amplitude of the modes at all scales blows up at the time corresponding to the divergence. At this point, the linear theory of cosmological perturbations that we used is no longer valid. One way to avoid this problem is to assume that all the physical wavenumber $k_{\rm phys} < M_{\rm PV}$ at the beginning of inflation. On the other hand, we also require that all the relevant perturbation modes are well inside the Hubble horizon, i.e., $k_{\rm phys} > H$ at the beginning of inflation, such that the quantum tensor perturbations originate from a Bunch-Davies vacuum state. With the above two assumptions, it is obvious that $\epsilon_* \ll 1$ during the slow-roll inflation. }

Similarly, the parameter $\mu_A$, which modifies the dispersion relation of the tensor modes, can be expressed in the form,
\bqn\lb{u}
\mu_A &=& \frac{\rho_A k (c_1 -c_2)/(a M_{\rm PV})}{1- \rho_A k c_1/(a M_{\rm PV})}\nb\\
&\simeq&-\rho_Ak\tau(c_1-c_2)\epsilon_*.
\eqn
It is worth noting that, in order to obtain the above expansions, we have used the relation
\bqn
a&=&-\frac{1}{\tau H}\left(1+\epsilon_1+\epsilon^2_1+\epsilon_1\epsilon_2\right)+\mathcal{O}(\epsilon^3).
\eqn

With the expressions of $z''/z$ and $\mu_A$, one observes that the equation of motion in Eq.(\ref{Eq}) can be casted into the form
\bqn\lb{zz}
&&u''_A+\Bigg\{\big[1-\rho_Ak\tau(c_1-c_2)\epsilon_*\big]k^2-\frac{v_t^2-\frac{1}{4}}{\tau^2} \nb\\
&&~~~ +\rho_A\frac{k}{\tau}c_1\epsilon_*\Bigg\}u_A=0.
\eqn
When $c_1=c_2$, this equation reduces to the same form as that in Chern-Simons gravity, which admits an exact solution in terms of confluent hypergeometric functions \cite{parity_power10}. However, when $c_1\neq c_2$,  this equation does not have exact solutions. In order to obtain its solution, we have to consider some approximations. In general, the most widely considered approach is the WKB approximation,  if the WKB condition is satisfied during the whole process. However, in some cases, the WKB condition may be violated or not be fulfilled completely (see \cite{zhu_inflationary_2014}). Recently, we have developed a mathematical approximation (the uniform asymptotic approximation) for better treatment to equations with turning points and poles, an approximation that has been verified to be powerful and robustness in calculating primordial spectra for various inflation models \cite{zhu_constructing_2014, zhu_inflationary_2014, zhu_quantum_2014, zhu_power_2014, martin_kinflationary_2013,ringeval_diracborninfeld_2010, wu_primordial_2017, qiao_inflationary_2018, ding_inflationary_2018, zhu_scalar_2015, zhu_detecting_2015, zhu_inflationary_2016, habib_inflationary_2002, habib_inflationary_2005, habib_characterizing_2004, zhu_inflationary_2014, zhu_highorder_2016, geng_schwinger_2018} and applications in studying the reheating process \cite{Zhu:2018smk} and quantum mechanics \cite{Zhu:2019bwj}.  In the following subsections, we apply this approximation to construct the approximate solution of (\ref{zz}) and calculate the primordial tensor power spectrum in the general ghost-free parity-violating gravity.

\subsection{Uniform asymptotic approximation}

In this subsection, we will apply the uniform asymptotic approximation method to construct the approximate asymptotic solutions. To proceed, let us first rewrite the equation of motion (\ref{zz}) in the following standard form \cite{zhu_constructing_2014, zhu_inflationary_2014},
\bqn
\frac{d^2u_A(y)}{dy^2}=[g(y)+q(y)]u_A(y),
\eqn
where $y\equiv-k\tau$ is a dimensionless variable and
\bqn
g(y)+q(y)\equiv\frac{v_t^2-\frac{1}{4}}{y^2}+\frac{\rho_Ac_1\epsilon_*}{y}-\rho_Ay(c_1-c_2)\epsilon_*-1.\nb\\
\eqn
In general, $g(y)$ and $q(y)$ have two poles (singularities): one is at $y=0^+$ and the other is at $y=+\infty$. Now, in order to construct the approximate solution in the uniform asymptotic approximation, one has to choose \cite{zhu_inflationary_2014, zhu_highorder_2016},
\bqn
q(y)&=&-\frac{1}{4y^2},
\eqn
to ensure the convergence of the errors of the approximate solutions around the second-order pole at $y=0^+$. With this choice, the function $g(y)$ is given by
\bqn
g(y)&=&\frac{v_t^2}{y^2}-1-\rho_Ay(c_1-c_2)\epsilon_*+\frac{\rho_Ac_1\epsilon_*}{y}.
\eqn
Except for the two poles at $y = 0^+$ and $y = +\infty$, $g(y)$ may also have a single zero in the range $y \in (0, +\infty)$, which called a single turning point of $g(y)$. By solving the equation $g(y) = 0$, we obtain the turning point,
\bqn
y_0^A&=&-\frac{1}{3\rho_A(c_1-c_2)\epsilon_*}[1-2^{1/3}(1+3\rho_A^2(c_1-c_2)c_1\epsilon^2_*)/Y\nb\\
&&-2^{-1/3}Y],
\eqn
where
\bqn
Y&=&(Y_1+\sqrt{-4(1+3\rho_A^2(c_1-c_2)c_1\epsilon^2_*)^3+Y_1^2})^{1/3},\nb\\
Y_1&=&-2+27v_T^2(\rho_Ac_1-\rho_Ac_2)^2\epsilon^2_*-9\rho_A^2(c_1-c_2)c_1\epsilon^2_*.\nb
\eqn
In the uniform asymptotic approximation, the approximate solution depends on the type of turning point. Thus, in the following discussion, we will discuss the solution around this turning point in details.

For the single turning point $y_0$, the approximate solution of equation of motion around this turning point can be expressed in terms of Airy type functions as \cite{zhu_highorder_2016, zhu_inflationary_2014}
\bqn\lb{ALY}
u_A=\alpha_0\left(\frac{\xi}{g(y)}\right)^{1/4}{\rm Ai}(\xi)+\beta_0\left(\frac{\xi}{g(y)}\right)^{1/4}{\rm Bi}(\xi),\nb\\
\eqn
where $\rm{Ai}(\xi)$ and $\rm{Bi}(\xi)$ are the Airy functions, $\alpha_0$ and $\beta_0$ are two integration constants, $\xi$ is the function of $y$ and the form of $\xi(y)$ is given by \cite{zhu_highorder_2016, zhu_inflationary_2014}
\bqn
\xi(y) =
\begin{cases}
\left(-\frac{3}{2}\int^y_{y_0}\sqrt{g(y')}dy'\right)^{2/3} ,\;  & y\leq y_0,\\
-\left(\frac{3}{2}\int^y_{y_0}\sqrt{-g(y')}dy'\right)^{2/3} ,\; & y\geq y_0.\\
\end{cases}\nb\\
\eqn
With this solution, we need to determine the coefficients $\alpha_0$ and $\beta_0$ by matching it with the initial condition in the limit $y \to +\infty$. For this purpose, we assume that the universe was initially in an adiabatic vacuum \cite{zhu_highorder_2016, zhu_inflationary_2014},
\bqn
\lim_{y \to +\infty} u_k(y)&=&\frac{1}{\sqrt{2\omega_k}}e^{-i\int\omega_k d\eta}\nb\\
&=&\sqrt{\frac{1}{2k}}\left(\frac{1}{-g}\right)^{1/4}\exp\left(-i\int^y_{y_i}\sqrt{-g}dy\right).\nb\\
\eqn
When $y\rightarrow+\infty$, we note that $\xi(y)$ is very large and negative. In this limit, the asymptotic form of the Ariy functions read
\bqn
{\rm Ai}(-x)&=&\frac{1}{\pi^{1/2}x^{1/4}}\cos\left(\frac{2}{3}x^{3/2}-\frac{\pi}{4}\right),\\
{\rm Bi}(-x)&=&-\frac{1}{\pi^{1/2}x^{1/4}}\sin\left(\frac{2}{3}x^{3/2}-\frac{\pi}{4}\right).
\eqn
Combining the initial condition and the approximate analytical solution, we obtain
\bqn
\alpha_0=\sqrt{\frac{\pi}{2k}}e^{i\frac{\pi}{4}},~~~
\beta_0=i\sqrt{\frac{\pi}{2k}}e^{i\frac{\pi}{4}}.
\eqn

\begin{figure*}
  \centering
  \includegraphics[width=8cm]{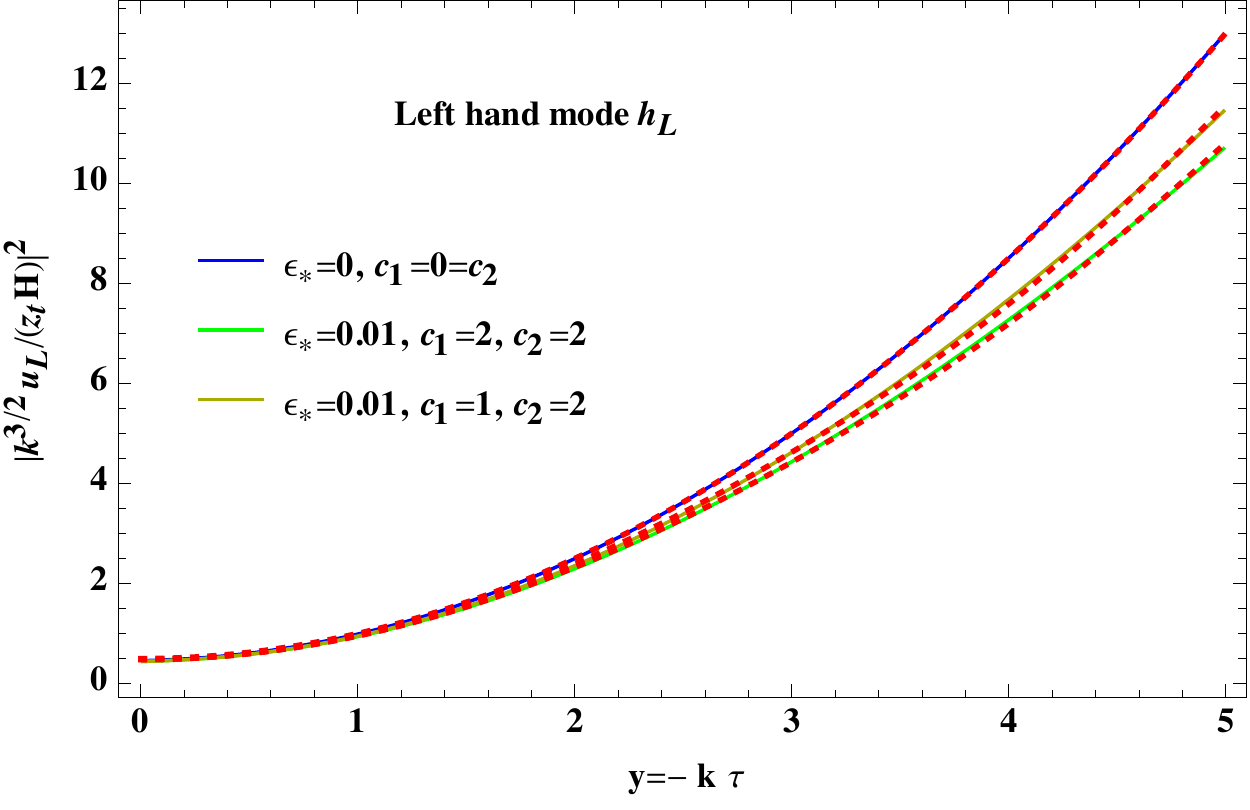}
    \includegraphics[width=8cm]{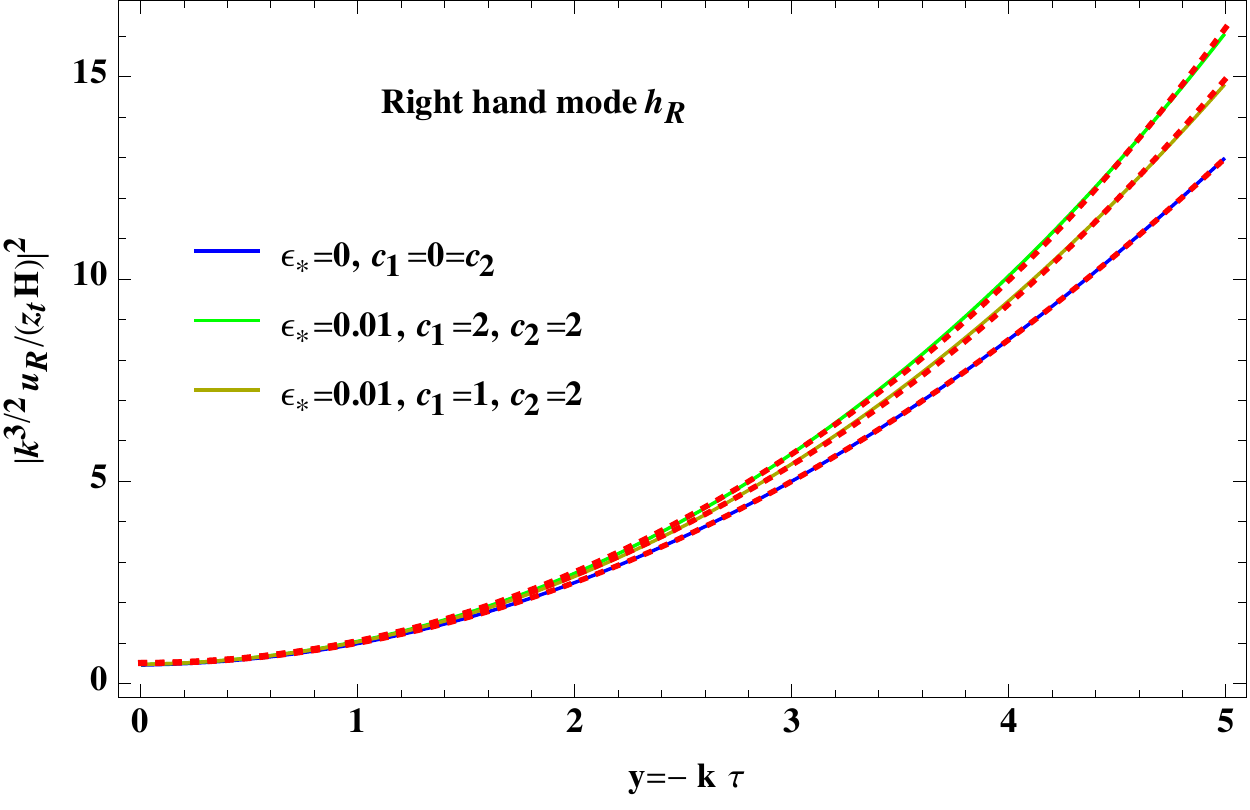}
  \caption{\red {The uniform asymptotic approximate solutions of mode functions $|k^{3/2}u_{A}/(z_t H)|^2$ (solid curves) and the corresponding numerical solutions (dotted curves). Left panel presents the solution of the left-hand mode while the right panel presents the right-hand mode. In each panel, the solid curves with blue, green, and darker yellow colors correspond to the solutions for cases of general relativity, Chern-Simons theory, and ghost-free parity-violating gravities respectively. The numerical solution associated with each analytical solutions are presented by the red dotted curves.}} \label{solution}
\end{figure*}

\red{Having obtained the approximate solutions of the mode functions $u_{\rm R, L}(y)$ as given by Eq.~(\ref{ALY}) with coefficients $\alpha_0$ and $\beta_0$ being given above, let us compare them with the numerical solutions. The results are presented in Fig.~\ref{solution}, in which we present the uniform asymptotic approximate solutions (solid blue, green, darker yellow curves) and numerical solutions (red dotted curves) of mode functions $|k^{3/2}u_{A}/(z_t H)|^2$ in general relativity, Chern-Simons theory, and ghost-free parity-violating gravities respectively. Left panel presents the solution of the left-hand mode while the right panel presents the right-hand mode. From this figure, one can see clearly that our analytical solutions are extremely close to the numerical ones, and even are not distinguishable from the numerical ones. For the  values of $c_1$, $c_2$, and $\epsilon_*$ we used in the figure, this figure also show that the right-hand mode trends to be enhanced while the left-hand mode trends to be suppressed by the parity violation. This feature is also consistent with the analytical results of the power spectra calculated later in the next subsection.
}

\subsection{Power spectra of PGWs}

Once the approximate solutions of the PGWs have been derived in the manner described above, the relevant
power spectra $\mathcal{P}^{L,R}_{\rm T}$ can be computed in the limit $y \to 0$ via
\bqn
\mathcal{P}_{\rm T}^{\rm L} = \frac{2 k^3}{\pi^2} \left|\frac{u_k^{\rm L}(y)}{z}\right|^2,
\mathcal{P}_{\rm T}^{\rm R} = \frac{2 k^3}{\pi^2} \left|\frac{u_k^{\rm R}(y)}{z}\right|^2.
\eqn
When $y\to 0^+$, the argument of Airy functions, $\xi(y)$, becomes very large and positive, allowing the use of the following asymptotic forms
\bqn\lb{AB}
{\rm Ai}(x)&=&\frac{1}{2\pi^{1/2}x^{1/4}}\exp\left(-\frac{2}{3}x^{3/2}\right),\\
{\rm Bi}(x)&=&-\frac{1}{\pi^{1/2}x^{1/4}}\exp\left(\frac{2}{3}x^{3/2}\right).
\eqn
From the Airy functions (\ref{AB}) we observe that, in this limit, only the growing mode of $u_k^A(y)$ is relevant, so we have
\bqn
u_k^A(y)&\approx&\beta_0\left(\frac{1}{\pi^2g(y)}\right)^{1/4}\exp\left(\int^{y_0}_ydy\sqrt{g(y)}\right)\nb\\
&=&i\frac{1}{\sqrt{2k}}\left(\frac{1}{g(y)}\right)^{1/4}\exp\left(\int^{y_0}_ydy\sqrt{g(y)}\right).\nb\\
\eqn
The power spectra of PGWs are then given by
\bqn\lb{PS}
\mathcal{P}^A_{\rm T}
&=&\frac{k^2}{\pi^2}\frac{1}{z^2}\frac{y}{v_t}\exp\left(2\int^{y_0}_ydy\sqrt{g(y)}\right)\nb\\
&\simeq&18\frac{H^2}{\pi^2 e^{3}}e^{\frac{\pi\rho_A\epsilon_*}{16}(9c_2-c_1)}\nb\\
&\simeq&18\frac{H^2}{\pi^2 e^{3}}\left[1+\frac{\pi\rho_A}{16}\mathcal{M}\epsilon_*+\frac{\pi^2 }{2\times16^2}\mathcal{M}^2\epsilon_*^2+\mathcal{O}(\epsilon_*)^3\right],\nb\\
\eqn
where we define a dimensionless parameter $\mathcal{M}  \equiv 9c_2-c_1$ and
\bqn
\frac{9c_2-c_1}{M_{\rm PV}}&=& 8\dot{\vartheta}+4 \dot{a_1}\dot{\phi}^2 +8 a_1\dot{\phi}\ddot{\phi} - 8a_1 H \dot{\phi}^2- 16\dot{a_2}\dot{\phi}^2 \nb\\
&&- 32a_2\dot{\phi}\ddot{\phi}-10\dot{a_3}\dot{\phi}^2 -20a_3\dot{\phi}\ddot{\phi} +4a_3 H \dot{\phi}^2 \nb\\
&&-32\dot{a_4}\dot{\phi}^2-64a_4\dot{\phi}\ddot{\phi}+2 b_1\dot{\phi}^3-4b_2\left(2 H\dot{\phi}^2-\dot{\phi}\ddot{\phi}\right)\nb\\
&&-2b_3\left(\dot{\phi}^3\ddot{\phi}- H \dot{\phi}^4\right)-2b_4\left(\dot{\phi}^3\ddot{\phi}- H \dot{\phi}^4\right)\nb\\
&&+2b_5 H \dot{\phi}^4-2b_7\dot{\phi}^3\ddot{\phi}.
\eqn
Obviously, the power spectra can be modified due to the presence of the parity-violating terms in the action (\ref{action}). As expected, one can check that, when $\mathcal{M}\epsilon_*= 0$, the standard GR result is recovered. Therefore, we can rewrite the power spectra in (\ref{PS}) as the following form,
\bqn
\mathcal{P}^A_{\rm T}= \frac{\mathcal{P}^{\rm GR}_{\rm T}}{2}\left[1+\frac{\pi\rho_A}{16}\mathcal{M}\epsilon_*+\frac{\pi^2\rho^2_A}{2\times16^2}\mathcal{M}^2\epsilon_*^2+\mathcal{O}(\epsilon_*)^2\right],\nb\\
\eqn
where
 \bqn
\mathcal{P}^{\rm GR}_{\rm T}=\frac{2 k^3}{\pi^2}\left(\left|\frac{u_k^{\rm L}(y)}{z}\right|^2+\left|\frac{u_k^{\rm R}(y)}{z}\right|^2\right)
\eqn
denotes the standard nearly scale invariant power-law spectrum calculated by uniform asymptotic approximation in the framework of GR \cite{zhu_inflationary_2014}. For the two circular polarization modes, i.e., $A={\rm R}$ and $A={\rm L}$, the spectra $\mathcal{P}^{\rm GR}_{\rm T}$ have the exactly same form. The quantity $\mathcal{M}$ depends on the coefficients $\vartheta$, $a_{\mathcal{A}}$ and $b_{\mathcal{A}}$, as well as the evolution of the scalar field. \red{It is interesting to observe that for positive value of $\mathcal{M}$, the parity violation trends to enhance (suppress) the power spectra of the right (left) hand modes.} During the slow-roll inflation, the scalar field is slow-rolling, which satisfies the slow-roll conditions (\ref{Con}).  With this condition, the quantities $c_1$ and $c_2$ are assumed to be slowly varying during the expansion of the universe, which can be approximately treated as constants during the slow-roll inflation. In the expression of $9c_2-c_1$, we observe that it contains the terms with $\vartheta, a_{\mathcal{A}}, b_{\mathcal{A}}$ and their derivatives with respect to $\phi$. Considering the scalar field $\phi$ with the slow-roll condition (\ref{Con}), the leading contribution to $9c_2-c_1$ reads
\bqn
\frac{9c_2-c_1}{M_{\rm PV}}\simeq 8 \dot \vartheta - 8 (a_1-\frac{a_3}{2}+ b_2) H \dot \phi^2.
\eqn
 Therefore, the leading contribution to the power spectrum of PGWs depends only on the coefficients $\dot \vartheta, a_1, a_3$ and $b_2$.

\subsection{The circular polarization and detectability}

Now, we are in a position to calculate the degree of the circular polarization of PGWs, which is defined by the differences of the amplitudes between the two circular polarization states of PGWs as
\bqn
\Pi&\equiv&\frac{\mathcal{P}^{\rm R}_{\rm T}-\mathcal{P}^{\rm L}_{\rm T}}{\mathcal{P}^{\rm R}_{\rm T}+\mathcal{P}^{\rm L}_{\rm T}} \simeq \red{ \frac{\pi}{16} (9c_2-c_1)\epsilon_* } +\mathcal{O}(\epsilon_*^3)\nb\\
&\simeq&\frac{\pi}{2} \dot \vartheta M_{\rm PV}\epsilon_*-\frac{\pi}{2}(a_1-\frac{a_3}{2}+ b_2)H \dot \phi^2M_{\rm PV} \epsilon_*+\mathcal{O}(\epsilon_*^3).\nb\\
\eqn
As expected, when $a_1 = a_3 = b_2 =0$, the above expression of the circular polarization $\Pi$ exactly reduces to that in Chern-Simons gravity \cite{Alexander_gravity_2008, Satoh_gravity_2005}. Obviously, under conditions (\ref{Con}), we observe that the degree of the circular polarization $\Pi$ is very small due to the suppressing parameter $\epsilon_*$.

As we mentioned in the introduction, the parity-violating effect in PGWs, which is measured by the observable $\Pi$, can produce the TB and EB spectra in the CMB data. This provides the opportunity to directly detect the chiral asymmetry of gravity by observations, which have been discussed in the literature (see Refs. \cite{parity_CMB, parity_CMB2, parity_CMB4} for examples). However, as pointed out in \cite{parity_power8}, the detectability of the circular polarization of PGWs is sensitive to the values of the tensor-scalar-ratio $r$ and $\Pi$. According to the combination of Planck 2018 data and the BICEP2/Keck Array BK15 data \cite{Akrami:2018odb}, $r$ has been tightly constrained as $r \lesssim 0.065$. For this case, in order to detect any signal of parity violation in the forthcoming CMB experiments,  $\Pi$ must be larger than $\mathcal{O}(0.5)$ as discussed, even in an ideal case with the cosmic variance limit. On the other hand, since the condition $\epsilon_* \ll 1$ is imposed for the considerations made in constructing the theory, the order of magnitude of $\Pi$ is roughly $\lesssim \mathcal{O}(0.5)$. For these reasons, we conclude that it seems difficult to detect or efficiently constrain the parity violation effects on the basis solely of two-point statistics from future CMB data.

\section{Conclusions and Outlook}
\renewcommand{\theequation}{5.\arabic{equation}} \setcounter{equation}{0}

In this paper, we study the circular polarization of PGWs in the ghost-free parity-violating theory of gravity, which generalizes the Chern-Simons gravity by including the first and second derivatives of the coupling scalar field. Applying the uniform asymptotic approximation to the equation of motion for the PGWs, we construct the approximate analytical solution to the PGWs during the slow-roll inflation. Using this approximate solution,  we calculate explicitly both the power spectra for the two polarization modes and the corresponding degree of circular polarization of PGWs. It is shown that with the presence of the parity violation, the power spectra of PGWs are slightly modified and the degree of circular polarization becomes nonzero. However, the circular polarization generated in the ghost-free parity-violating theory of gravity is quite small, which is suppressed by the energy scale of parity violation of the theory, and it is difficult to detect by using the power spectra of future CMB data.

It should be noted that in all the above discussions, the effect of the parity violation in the non-Gaussianity of PGWs has not been considered yet. Although there is very little hope to detect the parity-violating signatures in the two-point correlation of CMB data, a calculation in the Chern-Simons gravity shows that the parity-violating signatures in the bispectrum could be large enough and detectable in the future CMB data \cite{parity_power10} (Note that, similar analysis for the Horava-Lifshitz gravity with parity violation was also carried out in \cite{Zhu:2013fja}). According to the analysis in \cite{parity_power10}, the tensor-tensor-scalar bispectra for each polarization states can be peaked in the squeezed limit by setting the level of parity violation during inflation. Therefore, it is interesting to explore further whether the ghost-free parity-violating theory of gravity could lead to any parity-violating signatures in the non-Gaussianity of PGWs. We leave this topic as a separate work.


\section*{Acknowledgements}
A.W would like to express his gratitude to Zhejiang University of Technology for her hospitality, when part of the work was done. J.Q. and T.Z. are supported in part by NSFC Grants No. 11675143, the Zhejiang Provincial Natural Science Foundation of China under Grant No. LY20A050002, and the Fundamental Research Funds for the Provincial Universities of Zhejiang in China with Grant No. RF-A2019015. W.Z. is supported by NSFC Grants No. 11773028, No. 11633001, No. 11653002, No. 11421303, No. 11903030, the Fundamental Research Funds for the Central Universities, and the Strategic Priority Research Program of the Chinese Academy of Sciences Grant No. XDB23010200. A.W. is supported in part by NSFC Grants No. 11975203 and No. 11675145.

\end{document}